\documentclass[10pt,onecolumn, a4paper]{article}
\usepackage{lmodern}
\usepackage{amssymb,amsmath}
\usepackage{ifxetex,ifluatex}
\usepackage{fixltx2e} 
\ifnum 0\ifxetex 1\fi\ifluatex 1\fi=0 
  \usepackage[T1]{fontenc}
  \usepackage[utf8]{inputenc}
\else 
  \ifxetex
    \usepackage{mathspec}
  \else
    \usepackage{fontspec}
  \fi
  \defaultfontfeatures{Ligatures=TeX,Scale=MatchLowercase}
\fi
\IfFileExists{upquote.sty}{\usepackage{upquote}}{}
\IfFileExists{microtype.sty}{%
\usepackage{microtype}
\UseMicrotypeSet[protrusion]{basicmath} 
}{}
\usepackage[margin=1in]{geometry}
\usepackage{hyperref}
\hypersetup{unicode=true,
            pdftitle={Early warning signals for desynchronization in periodically forced systems},
            pdfauthor={Pablo Rodríguez-Sánchez; Egbert H. Van Nes; Marten Scheffer},
            pdfkeywords={quasipotential, visualization},
            pdfborder={0 0 0},
            breaklinks=true}
\usepackage{natbib}
\usepackage{longtable,booktabs}
\usepackage{graphicx,grffile}
\makeatletter
\def\maxwidth{\ifdim\Gin@nat@width>\linewidth\linewidth\else\Gin@nat@width\fi}
\def\maxheight{\ifdim\Gin@nat@height>\textheight\textheight\else\Gin@nat@height\fi}
\makeatother
\setkeys{Gin}{width=\maxwidth,height=\maxheight,keepaspectratio}
\IfFileExists{parskip.sty}{%
\usepackage{parskip}
}{
\setlength{\parindent}{0pt}
\setlength{\parskip}{6pt plus 2pt minus 1pt}
}
\ifx\paragraph\undefined\else
\let\oldparagraph\paragraph
\renewcommand{\paragraph}[1]{\oldparagraph{#1}\mbox{}}
\fi
\ifx\subparagraph\undefined\else
\let\oldsubparagraph\subparagraph
\renewcommand{\subparagraph}[1]{\oldsubparagraph{#1}\mbox{}}
\fi

\let\rmarkdownfootnote\footnote%
\def\footnote{\protect\rmarkdownfootnote}

\usepackage{titling}

  \title{Early warning signals for desynchronization in periodically forced
systems}
    \pretitle{\vspace{\droptitle}\centering\huge}
  \posttitle{\par}
    \author{Pablo Rodríguez-Sánchez \\ Egbert H. Van Nes \\ Marten Scheffer}
    \preauthor{\centering\large\emph}
  \postauthor{\par}
      \predate{\centering\large\emph}
  \postdate{\par}
    \date{2020-02-29}

\usepackage{tikz}
\usepackage{pgfplots}
\usepackage{float}
\usetikzlibrary{shapes,arrows}

\begin{document}
\maketitle
\begin{abstract}
Conditions such as insomnia, cardiac arrhythmia and jet-lag share a
common feature: they are all related to the ability of biological
systems to synchronize with the day-night cycle. When organisms lose
resilience, this ability of synchronizing can become weaker till they
eventually become desynchronized in a state of malfunctioning or
sickness. It would be useful to measure this loss of resilience before
the full desynchronization takes place. Several dynamical indicators of
resilience (DIORs) have been proposed to account for the loss of
resilience of a dynamical system. The performance of these indicators
depends on the underlying mechanism of the critical transition, usually
a saddle-node bifurcation. Before such bifurcation the recovery rate
from perturbations of the system becomes slower, a mechanism known as
critical slowing down. Here we show that, for a wide class of biological
systems, desynchronization happens through another bifurcation, namely
the saddle-node of cycles, for which critical slowing down cannot be
directly detected. Such a bifurcation represents a system transitioning
from synchronized (phase locked) to a desynchronized state, or vice
versa. We show that after an appropriate transformation we can also
detect this bifurcation using dynamical indicators of resilience. We
test this method with data generated by models of sleep-wake cycles.
\end{abstract}

\section{Introduction}\label{introduction}

The phenomenon of endogenous circadian rhythms, first observed by the
French polymath Jean-Jacques d'Ortous de Mairan in 1729
(\citet{DeMairan1729}), has transcended science to become part of the
popular culture, often referred to as \emph{the inner clock}. The
evolutionary convenience of synchronizing such \emph{inner clocks} with
the external cues, usually provided by regular astronomical events such
as day-night periods and seasons, is well established
(\citet{circadian2017}). Synchronization, thus, proves useful for living
systems and a difficulty to synchronize (and sometimes also to
desynchronize) can be an indicator of sickness or malfunctioning. Some
synchronization-related conditions include insomnia, jet-lag, arrhythmia
or epilepsy (\citet{Glass2001}).

The transition from a synchronized to a desynchronized regime is
discontinuous. The system is either synchronized or not. Therefore it
could be that synchronization is a special kind of critical transition
(\citet{Scheffer2009}). This is relevant as there have been developed
ways to foresee whether at critical transition is likely to occur
(\citet{Scheffer2009a}). These dynamic indicators of resilience (DIORs)
are based on the phenomenon of ``critical slowing down''
(\citet{Wissel1984}, \citet{VanNes2007}). According to this theory, the
recovery rate from perturbations decreases if systems are close to a
critical transition. In time series we can measure critical slowing down
using different indicators, such as increased autocorrelation and
variance (\citet{Dakos2012}).

In the present work we illustrate with simple models that some
transitions from synchronized to desynchronized states indeed can be
related to a special kind of critical transition, namely a saddle-node
bifurcation of cycles. We show that after an appropriate transformation
of the data, we can still use critical-slowing down indicators to see if
one of these transitions is likely to happen.

\section{Methods}\label{methods}

\subsection{Case study model}\label{sec:case-study-model}

Our goal is to develop generic indicators for the risk of
desynchronization of biological cycles such as the sleep-wake cycle. To
understand the properties of this system, we analyze a generic model of
such periodically forced cyclic systems. This minimal model consists of
two oscillators: a master (representing the external forcing, for
instance of a diurnal rhythm) and a slave (representing the organism's
state, for instance its sleep/awake status). We represent each
oscillator by its most basic feature: phase (\(\theta_\odot\) for the
master and \(\theta\) for the slave) The master's frequency is constant
(i.e.~the phase grows steadily from \(0\) to \(2\pi\) in \(24\) h), and
it is not affected by the slave's dynamics. The slave's dynamics are
more complex: in the absence of coupling it has a natural frequency, and
an increasing tendency to synchronize with the master if the coupling
gets more intense. These features are captured by model
\eqref{eq:kurcirgen}.

\begin{equation}
  \begin{cases}
    \frac{d \theta}{dt}       &= \omega - k \cdot f(\theta - \theta_\odot) \\
    \frac{d \theta_\odot}{dt} &= \omega_\odot \\
  \end{cases}
  \label{eq:kurcirgen}
\end{equation}

In model \eqref{eq:kurcirgen} each oscillator shows a natural frequency
(\(\omega\) and \(\omega_\odot\)). The first oscillator shows a tendency
to slow-down if \(\theta\) is ahead of \(\theta_\odot\), and to speed-up
otherwise. The function \(f\) measures the difference between \(\theta\)
and \(\theta_\odot\). Note that \(f\) has to be a periodic function (in
the sense of \(f(x+2\pi)=f(x)\)). This is a consequence of the cyclic
nature of phases: by definition phases \(\theta\) and \(\theta+2\pi\)
represent the same point in a cycle, and thus, the same physical
reality. In most applications \(f\) is also continuous and smooth. The
strength of the coupling is given by the positive constant parameter
\(k\). If the coupling is not strong enough (relative to the difference
in natural frequencies), synchronization doesn't happen.

The system \eqref{eq:kurcirgen} becomes simpler (and even analytically
tractable) if we use the phase difference
\(\phi(t) \equiv \theta(t) - \theta_{\odot}(t)\) as a new state
variable. With this change of state variable, the system takes the form
\eqref{eq:kurcirgensim}, where, for convenience, we made
\(\Omega \equiv \omega - \omega_{\odot}\).

\begin{equation}
  \frac{d \phi}{dt} = \Omega - k \cdot f(\phi)
  \label{eq:kurcirgensim}
\end{equation}

For the sake of clarity, we will use \(f(\phi) = \sin(\phi)\) in the
rest of this work. As we discuss in the online appendix, we can do this
without loss of generality. With this choice, our model becomes a simple
subcase of the classical Kuramoto model (see \citet{Kuramoto1975},
\citet{Strogatz2000}). Equating \eqref{eq:kurcirgensim} to zero, the
stable and unstable equilibria of our system are easily found to be
\(\phi^*_{s} = \Delta\) and \(\phi^*_{u} = \pi - \Delta\), where
\(\Delta \equiv \arcsin \frac{\Omega}{k}\). It is important to note
that, for those equilibria to exist, condition \eqref{eq:condition} should
be satisfied. Intuitively, this means that our system can only
synchronize cycles whose difference in natural frequencies (\(\Omega\))
have at most the same order of magnitude as the coupling term (\(k\)).

\begin{equation}
  \mid \frac{\Omega}{k} \mid \leq 1
  \label{eq:condition}
\end{equation}

When condition \eqref{eq:condition} is satisfied, the system
\eqref{eq:kurcirgensim} tends naturally to the stable solution. In this
case, the phase difference \(\phi\) is constant, so both oscillators
have the same frequency (\(\omega_\odot\)) and are thus synchronized.
If, on the contrary, condition \eqref{eq:condition} is not satisfied, the
phase difference \(\phi\) never stabilizes and consequently
synchronization is not possible.

\begin{figure}

{\centering \includegraphics[width=400px]{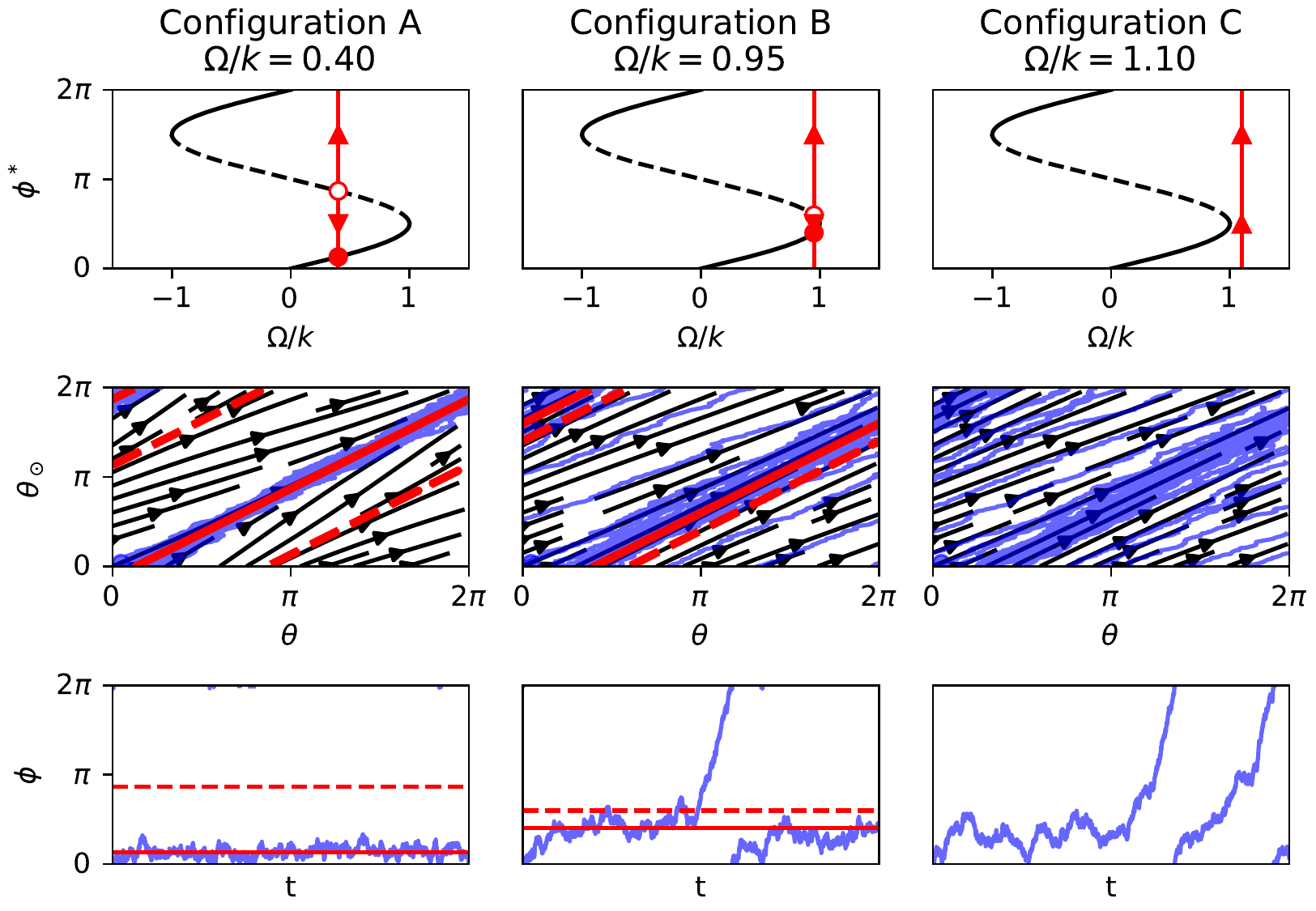} 

}

\caption{Each of the columns corresponds to a different
configuration for system \eqref{eq:kurcirgen}, identified by the value of
the bifurcation parameter \(\Omega / k\), which represents the
synchronization capacity. From left to right, each column represents
less coupling strength. Each of the rows corresponds to a different
representation of the dynamics. In the first row we see the bifurcation
diagram of the phase difference (\(\phi\)). The red arrows in the first
row represent the flow on the line. In the second and third rows, the
continuous red line represents the stable branch, and the dotted one the
unstable branch. Saddle-node bifurcations happen at
\(\frac{\Omega}{k} = \pm 1\). If \(\mid \frac{\Omega}{k} \mid > 1\) the
system has no equilibrium solution and, thus, represents a
desynchronized system. In both the second and third rows, the continuous
red lines represent the stable cycle, and the dashed line, the unstable
one. We plot in blue one simulated trajectory, under the influence of
stochastic noise (modelled as a Wiener process with a variance of
\(\sigma^2 = 0.04\)). The second row uses \((\theta, \theta_\odot)\) as
coordinates (phase space) and the third row uses \((t, \phi)\)
coordinates (time series). Notice that in the second column, even if
\(\frac{\Omega}{k} < 1\), a noise induced transition may happen due to
the proximity of the stable and unstable cycles.}\label{fig:plot-sto}
\end{figure}

But, what happens at the border between both cases, that is, when
\(\Omega / k\) approaches \(1\)? In such a situation, the stable and
unstable solutions collide and annihilate each other at
\(\phi^* = \frac{\pi}{2}\) (see first row of panel \ref{fig:plot-sto}).
This mechanism of losing stability is known as a saddle-node bifurcation
(\citet{Kuznetsov1998}, \citet{Strogatz1994}). For an extensive
discussion about the choice of this system, please refer to the
appendix.

We'll take advantage of the periodicity of our system by plotting its
trajectories over a \(2 \pi \times 2 \pi\) square with periodic boundary
conditions (or, equivalently, on the surface of a torus). When the phase
hits any border, it reappears at the opposite side (just like in
old-school video games such as \emph{Pac Man} or \emph{Asteroids}). In
figure \ref{fig:plot-sto} we see three different configurations of such
a system. As our parameter approaches the saddle-node bifurcation, both
the stable and unstable cycles get closer. When we introduce additive
noise to the dynamics, transitions can happen before the bifurcation is
reached if the noise is strong enough to make the state jump the gap
between both cycles (figure \ref{fig:plot-sto}, column B). Note that due
to the periodic boundaries the system is only momentarily desynchronized
as it ``collapses'' back to the synchronized dynamics.

\subsection{How to extract phase differences from
data?}\label{sec:link-with-data}

\begin{figure}

{\centering \includegraphics[width=500px]{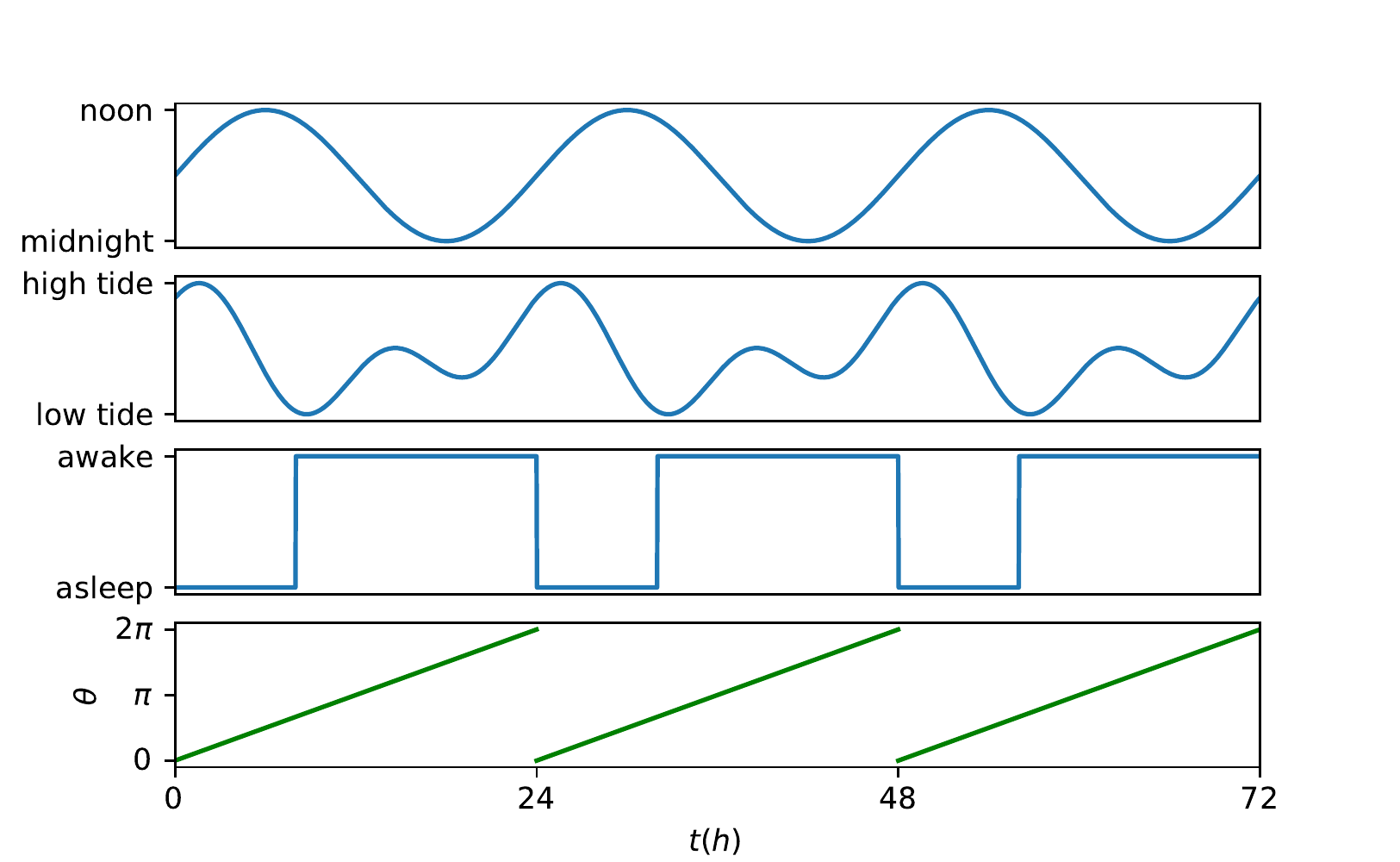} 

}

\caption{The first row shows the Sun's angular height from a local
horizon. Second row represents the height of the tide. Third row shows a
sleep wake cycle of a healthy individual. The fourth and last row shows
a common phase for the three above-mentioned phenomena (thus, the time
series in all rows can be expressed in the form given in equation
\eqref{eq:referenceState}). All series have been plotted for three whole
periods.}\label{fig:plot-obs}
\end{figure}

From the previous subsection it should be clear that synchronization is
related to a fold bifurcation that occurs in the phase difference of the
internal clock of a system respective to that of the forcing. This phase
difference is usually not directly measurable. Instead, experimental
data of periodic phenomena usually gives us indirect information about
the phase. The angle of the Sun respective to the local meridian, the
height of the tide or even the subjective feeling of sleepiness or
hunger along the day are obviously affected by the phase of the cycle
under study (see figure \ref{fig:plot-obs}). But, can we use these
indirect measurements to robustly infer the phase?

In order to answer this question, we will translate the ideas
illustrated in the previous paragraph and figure \ref{fig:plot-obs} to
mathematical language. Particularly, we'll assume, as a working
hypothesis, that there is a certain functional relationship \(M\)
between the phase of the cycle \(\theta(t)\) and our observations
\(y(t)\) (equation \eqref{eq:circadianState}). Due to the periodic nature
of our problem, we expect \(M\) to have a period of \(2\pi\).

\begin{equation}
  y(t) = M[\theta(t)]
  \label{eq:circadianState}
\end{equation}

We define a reference cycle \(y^{ref}(t)\) based in our knowledge about
the system under study. For instance, if we are studying sleep cycles
and \(y(t)\) represents the asleep state, a reasonable choice for
\(y^{ref}(t)\) could be \(y^{ref}(t) = 0\) (awake) if \(t\) is between
\(8\) and \(24\) h, and \(1\) (asleep) otherwise. Such a function
represents the idealized sleeping cycle of a healthy individual. We
assume the reference cycle to be the result of applying the unknown
function \(M\) to the phase of the external forcing \(\theta_\odot\)
(equation \eqref{eq:referenceState}).

\begin{equation}
  y^{ref}(t) = M[\theta_\odot(t)]
  \label{eq:referenceState}
\end{equation}

If our system is either synchronized or subject to slow variations in
its external conditions, we can consider the phase difference
(\(\phi(t) \equiv \theta(t) - \theta_\odot(t)\)) approximately constant
over a given time span \([t_a, t_b]\). It can be shown (see appendix
section \ref{subsec:detailed}) that under this circumstances we can
expect that \(y(t)\) is just shifted in time respective to
\(y^{ref}(t)\) by a certain time delay \(\lambda\) (equation
\eqref{eq:shift}).

\begin{equation}
  y(t) = y^{ref}(t + \lambda)
  \label{eq:shift}
\end{equation}

\begin{figure}

{\centering \includegraphics[width=500px]{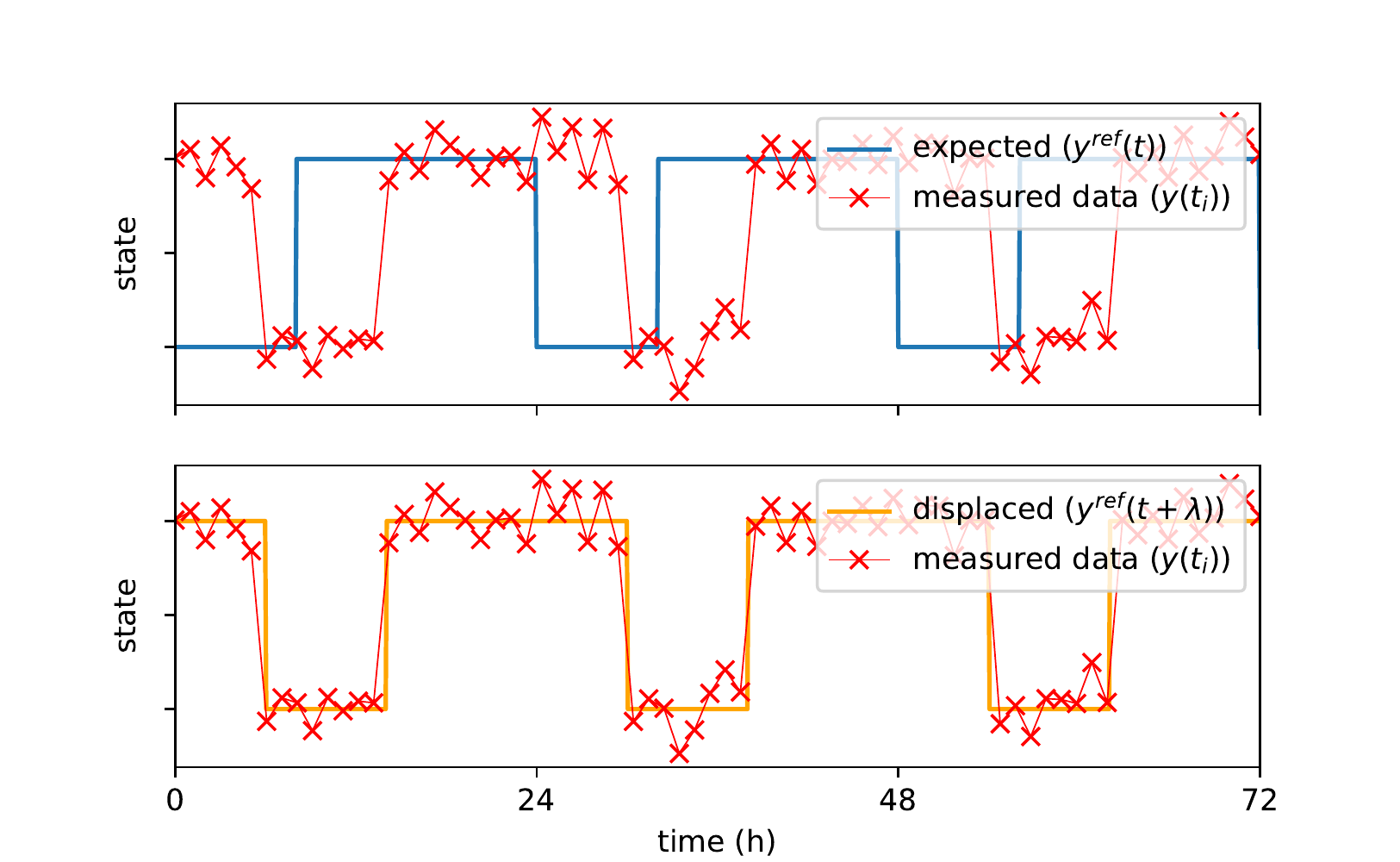} 

}

\caption{In both panels the red crosses represent the hypothetical
activity of a human being experiencing a jet lag, measured every 60
minutes during 3 days. In the upper row we see the expected daily
activity in blue (no activity while sleeping between \(0\) and \(8\)
hours, and activity the rest of the day). In the lower row we see, in
orange, the expected daily activity, but now displaced \(6h\) in the
time axis. This displacement provides the best fit for the data, and is
calculated by minimizing the function \(D^2(\lambda)\) given in equation
\eqref{eq:merit}.}\label{fig:merit}
\end{figure}

We find the time delay \(\lambda\) that best fits our data by minimizing
the sum of squares between the time-shifted reference cycle and our
measurements (see figure \ref{fig:merit} and equation \eqref{eq:merit}).
This time delay \(\lambda_{min}\) is proportional to the phase
difference. In section \ref{subsec:detailed} of the appendix, we show
that, specifically, \(\phi = \omega_\odot\lambda_{min}\).

\begin{equation}
  D^2(\lambda) = \sum_{i=a}^b (y_i - y^{ref}(t_i + \lambda))^2
  \label{eq:merit}
\end{equation}

Applying the method described above to different time windows allows us
to use data to estimate a time series of the phase differences
(\(\phi(t_j)\)) even if the precise analytical form of \(M\) is unknown
(see first row in figure \ref{fig:plot-synthetic-ews} and second row in
figure \ref{fig:philrob-plot}). See appendix section
\ref{subsec:detailed} for details.

\subsection{Resilience indicators}\label{sec:ews}

Saddle-node bifurcations are often preceded by the phenomenon of
critical slowing down (\citet{Scheffer2009a}). Such a phenomenon can be
directly observed in the time series even if the underlying dynamics are
unknown. In the present manuscript we used the above-mentioned
minimization algorithm along a moving window of typically 1-day width to
extract the phase difference. Then, we applied the methods proposed by
Dakos (\citet{Dakos2012}) to analyze this time series. Particularly, we
first detrended the time series by simply subtracting the average value
over non-intersecting windows of 1 day length. Afterwards, we calculated
the standard deviation and autocorrelation of the residuals, using a
rolling window with a length around 25-50\% of the original time series'
length. The optimal parameters (window length, autocorrelation lag,
etc.) depend on the time scale and characteristics of the data under
study. For more details about this method, see \citet{Dakos2012}. For an
extended discussion about the limitations of these methods, see
\citet{Dakos2015}.

\begin{figure}

{\centering \includegraphics[width=300px]{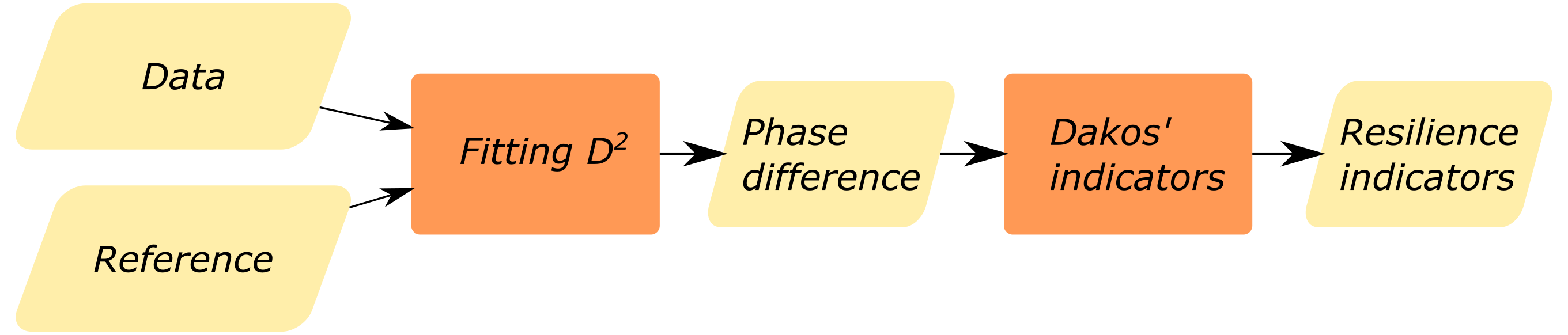} 

}

\caption{Schematic outline of our method.}\label{fig:flow}
\end{figure}

Figure \ref{fig:flow} summarizes all the steps, inputs and outputs of
our method.

\subsection{Model-generated time
series}\label{model-generated-time-series}

We tested our method with two model-generated time series.

The first time series was generated with the help of the sleep-wake
model of Strogatz (\citet{Strogatz1987}). We configured the system so
the time series represents the sleep-wake dynamics of an individual that
is becoming progressively more prone to insomnia. The insomnia effect
was simulated by allowing the coupling term \(k\) to linearly decrease
to zero along a period of \(135\) days. This makes the individual's
inner clock progressively less capable of coupling with the day-night
cycle, and eventually completely unable to do so. Strogatz's model
(\citet{Strogatz1987}) can be understood as a Kuramoto oscillator
followed by a postprocessing function \(M\) that transforms the inner
clock's phase \(\theta\) into a sleep-wake time series. Particularly,
\(M(\theta)\) returns \(1\) (awake) if the inner clock's phase
\(\theta\) is between \(2 \pi / 3\) and \(2 \pi\) radians (corresponding
to \(8\) and \(24\) hours in the inner clock), and \(0\) otherwise
(asleep). We used the generated time series to estimate the phase
difference. As this model contains an explicit phase, we can use it as a
control, and compare it with our estimated phase as a verification of
our method for extracting phases from data (see first row of figure
\ref{fig:plot-synthetic-ews}).

To show the generality of our method, we applied it to a second time
series generated with a more realistic model, the Phillips-Robinson
model (\citet{Phillips2007}). The Phillips-Robinson model is a
deterministic sleep-wake model based on neurological considerations, and
it doesn't contain an explicit phase. It describes the time evolution of
three state variables: \(V_v\) the activity of the ventrolateral
preoptic area (prompting the body to stay asleep), \(V_m\) the activity
of the mono aminergic group (prompting the body to stay awake) and \(H\)
the homeostatic pressure (an auxiliary variable that quantifies the need
for sleep). The dynamics of the model are given by the equations
\eqref{eq:philrob}, where \(F(V)\) is a saturation function given by
\eqref{eq:saturation} and \(C(t)\) (defined in equation \eqref{eq:forcing})
is a time-dependent external forcing, representing the astronomical
light/dark cycle. The remaining elements in equation \eqref{eq:philrob},
including the influence of the acetylcholine group (\(V_{a0}\)), are
just constants. The parameters used are the same as in
(\citet{Phillips2007}); see section \ref{subsec:table} in the online
appendix. Additionally, this appendix section provides a graphical
representation of the relationships in equation \eqref{eq:philrob}.

\begin{equation}
  \begin{cases}
  \begin{tabular}{r c r r r r r r}
  $\tau_v \frac{d V_v}{dt}$ & $=$ & $-V_v$ &  $               $ & $-\nu_{vm}S(V_m)$ & $+\nu_{vh}H$ & $-\nu_{vc}C(t)     $\\
  $\tau_m \frac{d V_m}{dt}$ & $=$ & $-V_m$ &  $-\nu_{mv}S(V_v)$ & $               $ & $          $ & $+\nu_{ma}S(V_{a0})$\\
  $\chi \frac{d H}{dt}    $ & $=$ & $  -H$ &  $               $ & $+\mu S(V_m)    $ & $          $ & $                  $
  \end{tabular}
  \end{cases}
  \label{eq:philrob}
\end{equation}

\begin{equation}
  S(V) = \frac{Q_{max}}{1 + e^{-\frac{V-\theta}{\sigma}}}
  \label{eq:saturation}
\end{equation}

\begin{equation}
  C(t) = \frac{1}{2} \left( 1 + \cos(\omega t + \alpha) \right)
  \label{eq:forcing}
\end{equation}

Once again, we simulated an individual whose sleep quality is slowly
deteriorating. We achieved this effect by allowing the coupling
parameter \(\nu_{vc}\) to decrease linearly from its normal value of
\(6.3\) \(mV\) to \(0\) \(mV\) along a period of three months. By doing
this, the ability of the subject to synchronize his internal clock with
the external time cues slowly disappears. The first episode of
insomnia/desynchronization happens on the 83rd day (see first row in
figure \ref{fig:philrob-plot}).

In order to simulate the fluctuations expected in any biological system
we added noise to the integration of both our time series. In
particular, we modeled our systems as Wiener processes. The
deterministic terms have been described in the previous paragraphs. The
stochastic terms (\(dW = \sigma dt\)) for Strogatz's model where set to
\(\sigma = 0.05\) for the inner clock's phase, and to \(0\) for the
driver. For the Phillips-Robinson model, the stochastic term was set to
\(\sigma = 1\) for the states \(V_v\) and \(V_m\), and \(0\) for \(H\).
The integration was performed numerically with the Python package
\emph{sdeint}.

\section{Application}\label{sec:results}

\begin{figure}

{\centering \includegraphics[width=500px]{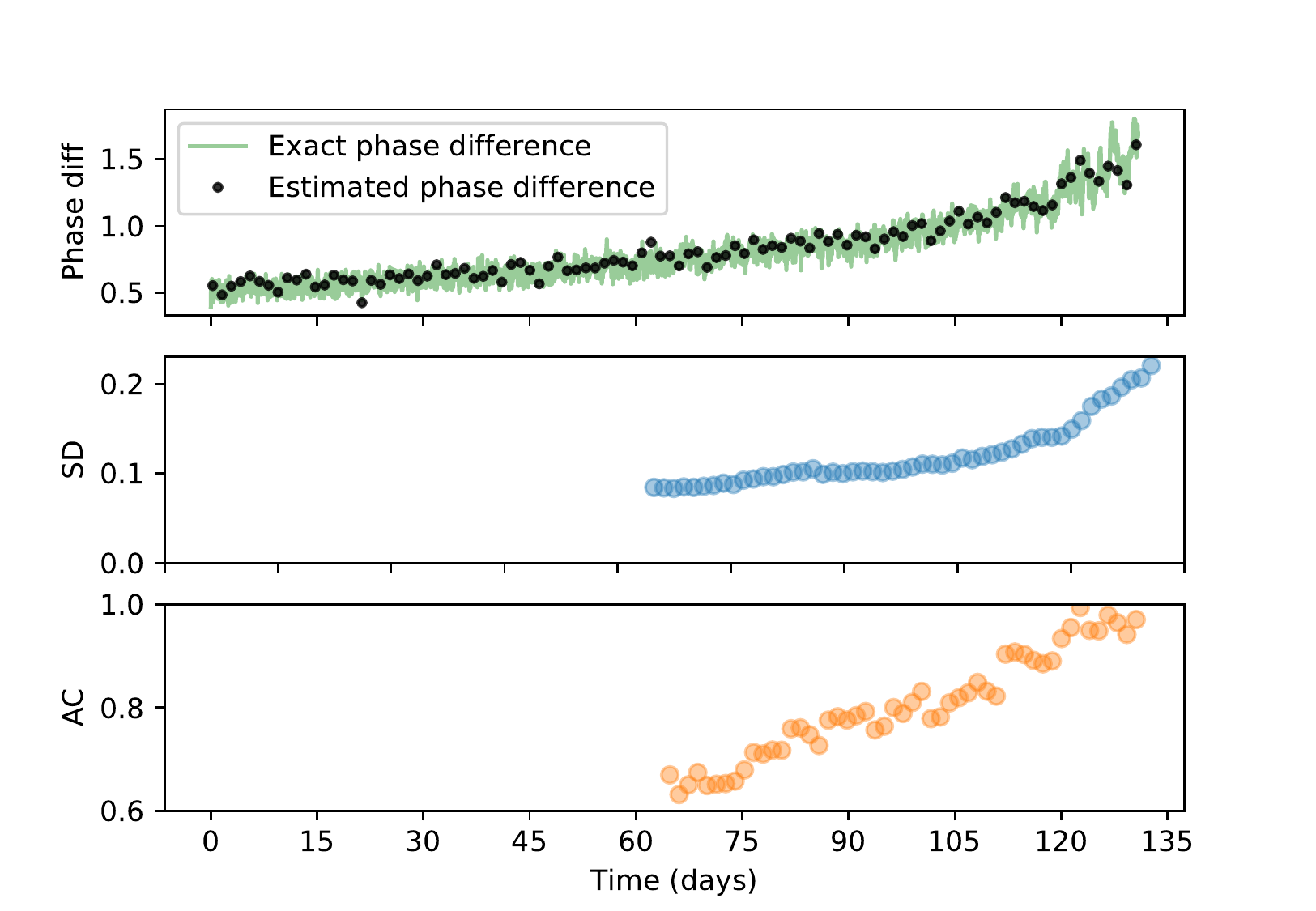} 

}

\caption{The black dots in the upper panel represent the
phase difference, as approximated by our method. The exact phase
difference is also shown (green line) as a reference of the method's
accuracy. The central and lower panel show the standard deviation (in
blue) and the autocorrelation with a \(24\) h lag (in orange) of the
estimated phase difference, both of them calculated for a window 50\%
the length of the data. In both panels, as the time increases, the
resilience of our system gets weaker.}\label{fig:plot-synthetic-ews}
\end{figure}

We applied the minimization algorithm described in the methods section
to the sleep-awake time series generated with Strogatz's model (see
methods). As a reference of a healthy sleep-wake cycle, we used a simple
assumption: a healthy individual is awake between \(8\) in the morning
and \(24\) at night, and asleep otherwise. We managed to reconstruct
correctly the phase difference. The reconstructed phase difference shows
the classical signs of slowing down (namely, increase in standard
deviation and autocorrelation) when the system is approaching the
bifurcation (see figure \ref{fig:plot-synthetic-ews}).

\begin{figure}

{\centering \includegraphics[width=500px]{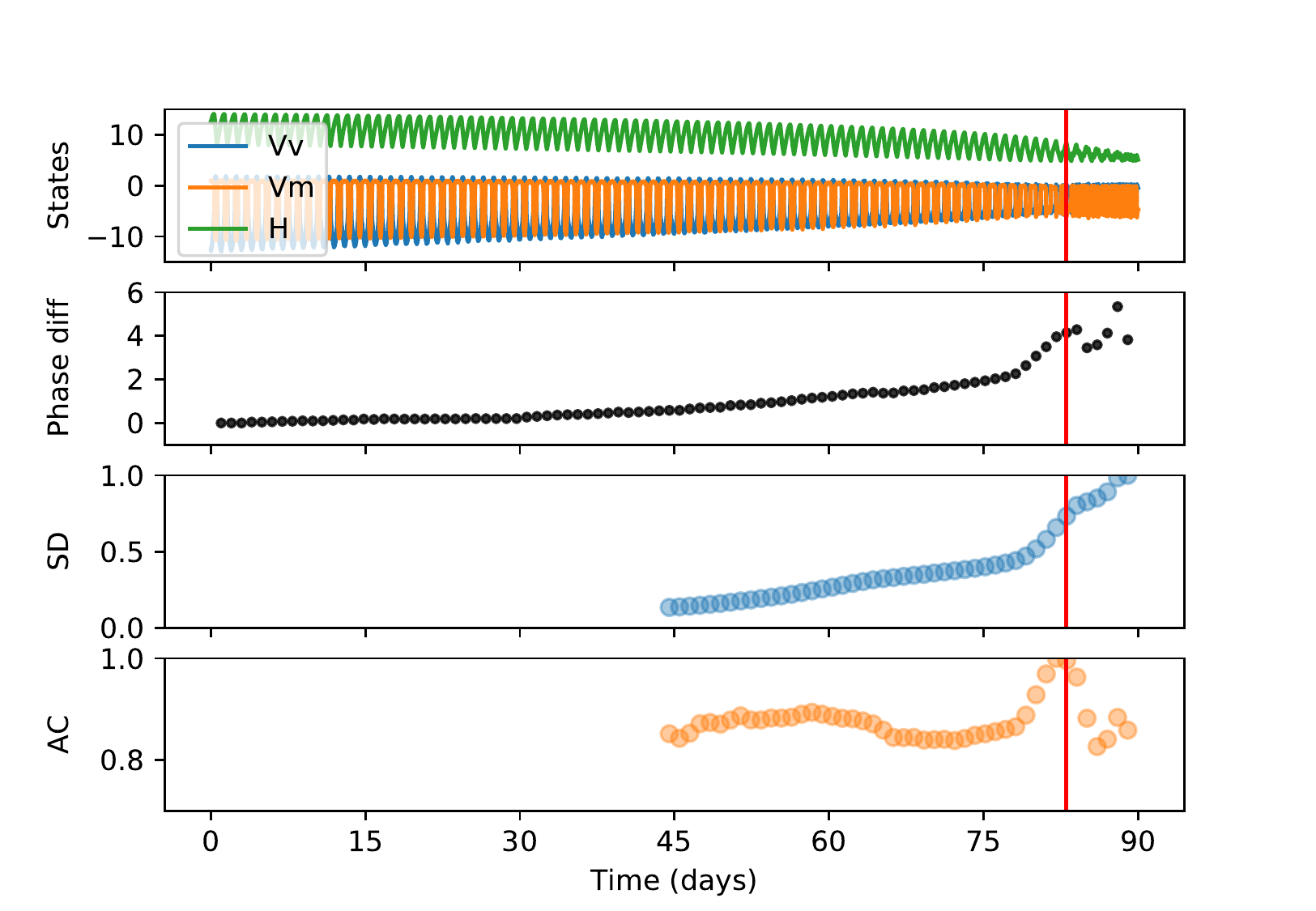} 

}

\caption{The upper panel shows a simulated time series
obtained by integrating the Phillips-Robinson model under the influence
of stochastic noise, and with the parameter \(\nu_{vc}\) decreasing
linearly in time to simulate an increasing difficulty in synchronizing.
The simulation was initialized with non-transient values, to ensure the
first days represent a healthy sleep wake cycle. The second panel shows
the estimated phase difference (in hours) using the method described in
our paper (sub-sampling the whole time series once per day, and using
day 1 as reference). Our reference time series was chosen to be the
somnogen level \(H\) during the first day. The two lower panels show the
standard deviation (in blue) and the autocorrelation with a \(48\) h lag
(in orange), calculated for a window of 45 days into the past, of the
estimated phase differences. The red line the 83rd day marks the first
episode of insomnia/desynchronization.}\label{fig:philrob-plot}
\end{figure}

Our method was also applied with success to the time series generated
with the Phillips-Robinson model (equation \eqref{eq:philrob}). We focused
our attention only in the time series corresponding to the state
variable \(H\). We used the time series corresponding to day \(1\) as a
reference to estimate the phase difference. Particularly, we built
\(y^{ref}(t)\) as a quadratic interpolator of the measurements
corresponding to the first day. As we can see in figure
\ref{fig:philrob-plot}, the first episode of insomnia (83rd day) is
preceded by an increase in both standard deviation and autocorrelation
of the estimated phase difference.

\section{Discussion}\label{discussion}

In the current work we presented a way of deriving dynamic indicators of
resilience (DIORs) for systems transitioning from synchronized to
desynchronized states through the family of bifurcations known as
saddle-node of cycles. Our method is designed for time series, and
doesn't require detailed knowledge of the deterministic dynamics of the
system. This makes it particularly suitable for biological systems where
a loss of synchronization may have an undesired effect (such as insomnia
or arrhythmia (\citet{Glass2001})) or may be an indicator of a loss of
resilience (such as the disruption in daily activity patterns in cows
after calving (\citet{VanDixhoorn2018})).

It may be argued that our method rests on the particular choice of the
model given in equation \eqref{eq:kurcirgen}. As we discuss in the
appendix, equation \eqref{eq:kurcirgen} represents the simplest, albeit
non-trivial representative of a broader family of synchronization
dynamics. Different choices yield different geometries in the
bifurcation diagram (figure \ref{fig:plot-sto}, first row), but the main
characteristic, the fact that at least one saddle-node bifurcation
exists, remains true. This, together with the method to extract phase
differences for general time series of periodically forced systems,
makes our approach valid under very general circumstances. Two
application examples of time series that were generated with two
different sleep-wake models, Strogatz's and Phillips-Robinson's are
analyzed.

The method to extract phase differences requires an approximate
reference time series. In the Strogatz's model application example we
used the very simple assumption that a healthy individual sleeps from 0
h at night to 8 h in the morning. Some problems may benefit from or even
require more sophisticated assumptions. In the absence of any detailed
knowledge of the system under study, another approach could be using the
dynamics of an arbitrary day as a reference. This is what we did in the
Phillips-Robinson application example.

Additionally, our method requires high quality time series. Those time
series should be long (as we need many cycles to infer the indicators)
and should have a high density of data points (typically of the order of
\(10\) points measured per cycle, depending on the shape of the time
series). This makes our method less suitable to be applied with success
in fields where the data is difficult and/or expensive to collect.
Luckily, data-rich systems such as the ones provided by wearable devices
are becoming increasingly popular in medicine or veterinary sciences.

Even after extracting the phase difference, the critical slowing down
may be difficult to detect under some circumstances. As already noted in
\citet{Dakos2015}, his method to forecast saddle-node bifurcations has
some fundamental limitations. For instance, it is required that the
time-scale of the changes in the external forcing to be slower than the
natural time-scale of the system (that is, the transitions shouldn't be
too sudden). The role of noise is also a delicate issue. On one hand,
noise is required in order to observe the phenomenon of critical slowing
down in the vicinity of a saddle-node bifurcation. On the other hand, it
obscures the deterministic dynamics. Our analysis will prove weak for
systems whose dynamics are strongly dominated by noise. Our method,
based on Dakos' indicators, shares this set of limitations.

Even with those applicability challenges, we consider our method to be a
step in the direction of forecasting transitions between synchronized
and desynchronized states. The fact that the disruption in certain
physiological rhythms is associated with disease (\citet{Glass2001}),
together with the recent increase in the availability of high-quality
biometric time series, makes the analysis and potential forecasting of
these relevant kind of transitions a topic worth being explored.

\section{Acknowledgments}\label{acknowledgments}

We thank Ingrid van Dixhoorn and Rudi de Mol for their useful comments
and suggestions.

This work was supported by funding from the European Union's
\textit{Horizon 2020} research and innovation programme for the
\textit{ITN CRITICS} under Grant Agreement Number 643073.

\newpage

\section*{Appendix}\label{appendix}
\addcontentsline{toc}{section}{Appendix}

\setcounter{equation}{0} \setcounter{figure}{0} \setcounter{section}{0}
\setcounter{subsection}{0}
\renewcommand{\theequation}{A.\arabic{equation}}
\renewcommand{\thefigure}{A.\arabic{figure}}
\renewcommand{\thesubsection}{A.\arabic{subsection}}

\subsection{Detailed derivation of the phase extracting
method}\label{subsec:detailed}

In order to justify the results of section \ref{sec:link-with-data}, we
will make use of equations \eqref{eq:kurcirgen}, \eqref{eq:circadianState},
\eqref{eq:referenceState} and \eqref{eq:shift}. As a first step, we will
write both sides of equation \eqref{eq:shift} in terms of \(M\). We can do
this by directly applying \eqref{eq:circadianState} and
\eqref{eq:referenceState}. The result is shown in equation
\eqref{eq:intermediate}.

\begin{equation}
  \begin{cases}
    y(t)                 &= M[\theta(t)] \\
    y^{ref}(t + \lambda) &= M[\theta_\odot(t + \lambda)] \\
  \end{cases}
  \label{eq:intermediate}
\end{equation}

In order be able to compare the functions given by equations
\eqref{eq:circadianState} and \eqref{eq:referenceState} we will change their
coordinates. Using the phase difference
(\(\phi(t) \equiv \theta(t) - \theta_\odot(t)\)), we can rewrite the
equation \eqref{eq:circadianState} in terms of \(\theta_\odot\) and
\(\phi\) (see first line in equation \eqref{eq:comparer}). Introducing a
leftwards shift \(\lambda\) in the time coordinate of equation
\eqref{eq:referenceState} it takes the form
\(y^{ref}(t + \lambda) = M[\theta_\odot(t + \lambda)]\), where
\(\theta_\odot(t + \lambda)\) can be evaluated exactly by its first
order Taylor expansion, that is,
\(\theta_\odot(t + \lambda) = \theta_\odot(t) + \omega_\odot \lambda\)
(cf.~second line of equation \eqref{eq:kurcirgen}). The results of both
coordinate transformations appear in equation \eqref{eq:comparer}.

\begin{equation}
  \begin{cases}
    y(t)                 &= M[\theta_\odot(t) + \phi(t)] \\
    y^{ref}(t + \lambda) &= M[\theta_\odot(t) + \omega_\odot \lambda] \\
  \end{cases}
  \label{eq:comparer}
\end{equation}

Note that if the system is synchronized and/or if the adiabatic
approximation (that is, that the external conditions vary slowly) holds
in our region of interest (\(t \in [t_a, t_b]\)), \(\phi(t)\) can be
approximated by a constant \(\phi\). We can estimate the value of
\(\phi\) by finding the shift \(\lambda^{min}\) that minimizes the
square distance between both functions (equation \eqref{eq:integral}). By
direct inspection of equation \eqref{eq:comparer} we see that this optimal
value corresponds to a phase difference of
\(\phi = \omega_\odot \lambda^{min}\).

\begin{equation}
  D^2(\lambda) = \int^{t_b}_{t_a} (y(s) - y^{ref}(s + \lambda))^2 ds
  \label{eq:integral}
\end{equation}

When faced with experimental data, we'll have a collection of \(N\)
measured values \(y_i\) sampled at times \(t_i\) (that is,
\(y_i = y(t_i)\)). The discrete equivalent of equation
\eqref{eq:integral}, representing the square distance between our measured
and the expected points, is given in \eqref{eq:merit}. By finding the
value of the time displacement \(\lambda\) that minimizes
\(D^2(\lambda)\), we find the time delay that better fits our data (see
figure \ref{fig:merit}).

\subsection{Further generalization}\label{subsec:generalization}

By manipulating the parameter \(\Omega / k\) in an equation like
\eqref{eq:kurcirgensim} with any non trivial continuous function
\(f(\phi)\) we are sure of encountering at least one saddle-node
bifurcation. Even more, the saddle-node is the only kind of bifurcation
that may happen.

\begin{figure}
\centering
\includegraphics{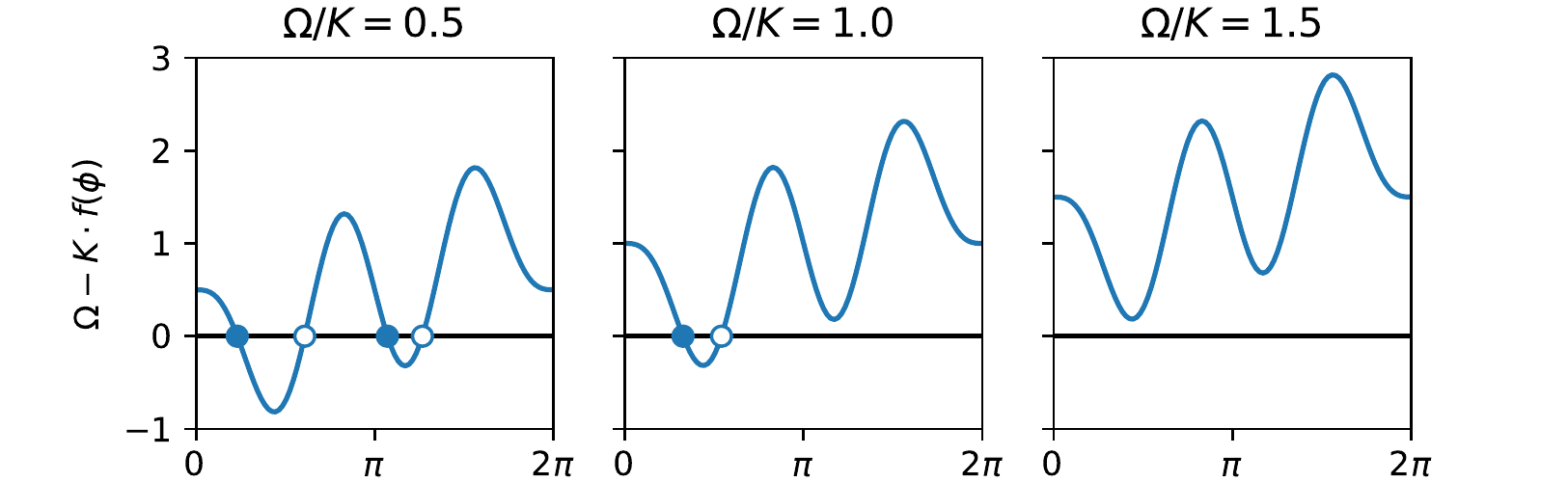}
\caption{\label{fig:plot-derivative-general}Here we plot the curve
\(y(\phi) = \Omega - k f(\phi)\) for a non-sinusoidal coupling function
\(f\). The function is non-constant, continuous, smooth and periodic. We
plotted it for three different values of the bifurcation parameter
\(\frac{\Omega}{k}\). The roots of each curve represent the equilibria
(filled dots if stable, white if unstable).}
\end{figure}

This can be proven graphically. As we discussed in the methods section,
the coupling function \(f\) in the model given by equation
\eqref{eq:kurcirgensim} can be any non-constant, continuous, smooth and
periodic function, not necessarily a sine. A function \(f\) satisfying
these properties will have at least one local minimum and one local
maximum per period. This is also true for the right-hand side of
equation \eqref{eq:kurcirgensim}. The effect of the parameter
\(\Omega / k\) is to move up and down the curve defined by
\(y(\phi) = \Omega - k f(\phi)\), whose roots represent the equilibria.
This rules out the pitchfork and the transcritical bifurcations, as
those require a change in the shape of the curve \(y(\phi)\)
(\citet{Strogatz2003}). By manipulating the parameter \(\Omega / k\),
the only possible bifurcations are collisions of stable and unstable
equilibria, that is, saddle-node bifurcations (\citet{Strogatz2003}).
Those bifurcations happen when a minimum or a maximum equals \(0\) (see
figure \ref{fig:plot-derivative-general}).

Those readers familiar with analysis may prefer noticing that, in the
vicinity of a minimum/maximum (\(\phi_0\)), the second order Taylor
expansion of the right-hand side of equation \eqref{eq:kurcirgensim} can
be written as the equation of a parabola \eqref{eq:parabola}.

\begin{equation}
  \Omega - k f(\phi) \approx \Omega - k f(\phi_0) + \frac{k f''(\phi_0)}{2} (\phi - \phi_0)^2
  \label{eq:parabola}
\end{equation}

By using the new variable
\(x \equiv \sqrt{\frac{k f''(\phi_0)}{2}}(\phi - \phi_0)\) (representing
a shift and re-scale of the horizontal axis), and renaming
\(\Omega - k f(\phi_0)\) as \(r\), the right-hand side of equation
\eqref{eq:parabola} adopts the canonical form of saddle-node bifurcation,
i.e.: \(r + x^2\) (\citet{Kuznetsov1998}).

Due to the generality of the conditions requested to the coupling
function \(f\), we expect saddle-node bifurcations in the phase
difference to be a widespread mechanism of synchronization and
desynchronization. Consequently, we expect those bifurcations to be
susceptible of being detected by the method described in this
manuscript.

\subsection{Parameters for Phillips-Robinson model}\label{subsec:table}

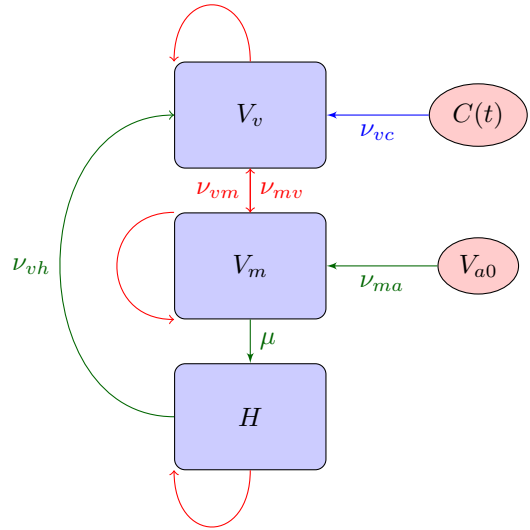
\begin{figure}
\centering
\begin{tikzpicture}[node distance = 2cm, auto]

    \tikzstyle{state} = [rectangle,draw,fill=blue!20,text width=5em,text centered,rounded corners,minimum height=4em]
    \tikzstyle{line} = [draw, -latex']
    \tikzstyle{source} = [draw, ellipse,fill=red!20, node distance=3cm,minimum height=2em]
    \definecolor{darkgreen}{rgb}{0,0.4,0}

    \node [state] (Vv) {$V_v$};
    \node [state, below of=Vv] (Vm) {$V_m$};
    \node [state, below of=Vm] (H) {$H$};

    \node [source, right of=Vv] (C) {$C(t)$};
    \node [source, right of=Vm] (A) {$V_{a0}$};

    \path [blue, line] (C) -> node{$\nu_{vc}$} (Vv);
    \path [darkgreen,line] (A) -> node{$\nu_{ma}$} (Vm);

    \path [darkgreen,line] (Vm) -> node{$\mu$} (H);
    \draw[red, ->] (Vv) -- node[midway, right] {$\nu_{mv}$} (Vm);
    \draw[red, ->] (Vm) -- node[midway, left] {$\nu_{vm}$} (Vv);
    \draw[darkgreen,->] (H.west) .. controls +(left:20mm) and +(left:20mm) .. node{$\nu_{vh}$} (Vv.west);

    \draw[red, ->] (Vv.north) .. controls +(up:10mm) and +(up:10mm) .. (Vv.north west);
    \draw[red, ->] (Vm.north west) .. controls +(left:10mm) and +(left:10mm) .. (Vm.south west);
    \draw[red, ->] (H.south) .. controls +(down:10mm) and +(down:10mm) .. (H.south west);

\end{tikzpicture}
\caption{Schematic summary of the dynamics of the Phillips-Robinson model. The light blue nodes represent the system's states ($V_v$ the activity of the ventrolateral preoptic area, $V_m$ the activity of the mono aminergic group and $H$ the homeostatic pressure). The pink nodes represent the external sources ($C(t)$, the astronomical light/dark forcing, and $V_{a0}$, the acetylcholine group constant influence). The positive effects are coded as green arrows. Negative ones as red arrows. Blue arrows represent oscillating effects.} \label{fig:diagram}
\end{figure}

The parameters used in equation \eqref{eq:philrob} are the same as in
(\citet{Phillips2007}), with the exception of \(\nu_{vc}\), that
decreases linearly from \(6.3\) mV to \(0\) mV along the period of
\(90\) days. The time units have been changed to hours. The dynamics of
the model are summarized in figure \ref{fig:diagram}. A usable
implementation of this model (in R) can be found at
\url{https://github.com/PabRod/sleepR}.

\begin{longtable}[]{@{}lcl@{}}
\toprule
Symbol & Value & Units\tabularnewline
\midrule
\endhead
\(\tau_m\) & \(10/3600\) & \(h\)\tabularnewline
\(\tau_v\) & \(10/3600\) & \(h\)\tabularnewline
\(\chi\) & \(10.8\) & \(h\)\tabularnewline
\(\nu_{vm}\) & \(1.9/3600\) & \(mV \cdot h\)\tabularnewline
\(\nu_{mv}\) & \(1.9/3600\) & \(mV \cdot h\)\tabularnewline
\(\nu_{vh}\) & \(0.19\) & \(mV \cdot nM^{-1}\)\tabularnewline
\(\mu\) & \(10^{-3}\) & \(nM \cdot h\)\tabularnewline
\(\nu_{vc}\) & \(6.3 - 0\) & \(mV\)\tabularnewline
\(\nu_{ma}S(V_{a0})\) & \(1\) & \(mV\)\tabularnewline
\(Q_{max}\) & \(100 \cdot 3600\) & \(h^{-1}\)\tabularnewline
\(\theta\) & \(10\) & \(mV\)\tabularnewline
\(\sigma\) & \(3\) & \(mV\)\tabularnewline
\(\omega\) & \(2 \pi / 24\) & \(h^{-1}\)\tabularnewline
\(\alpha\) & \(0\) & \(1\)\tabularnewline
\bottomrule
\end{longtable}

\newpage
\renewcommand\refname{References}
\bibliography{sync.bib}

\end{document}